\documentclass[10pt,journal,a4paper]{IEEEtran}
\usepackage{amssymb,amsmath}
\usepackage{cuted}
\usepackage{algorithm}
\usepackage{scalerel}
\usepackage{dsfont}
\usepackage[table]{xcolor}
\usepackage{gensymb}
\usepackage{mathbbol}
\usepackage{bbm}
\usepackage{algpseudocode}
\usepackage{cite}
\usepackage{graphicx,subfigure}
\usepackage{multirow}
\usepackage{psfrag}
\usepackage{url}
\usepackage{xcolor}
\usepackage[absolute,overlay]{textpos}
\usepackage{algpseudocode}
\usepackage{float}
\usepackage{lettrine}
\usepackage{indentfirst}
\usepackage{amsthm}

\newtheorem{theorem}{Theorem}

\usepackage{array}
\usepackage{psfrag}
\usepackage{url}
\usepackage{caption}
\usepackage[caption=false]{subfig}
\usepackage[font=footnotesize]{caption}
\usepackage{subfigure}

\makeatletter
\def\thickhline{%
  \noalign{\ifnum0=`}\fi\hrule \@height \thickarrayrulewidth \futurelet
   \reserved@a\@xthickhline}
\def\@xthickhline{\ifx\reserved@a\thickhline
               \vskip\doublerulesep
               \vskip-\thickarrayrulewidth
             \fi
      \ifnum0=`{\fi}}
\makeatother

\newlength{\thickarrayrulewidth}
\begin{document}
	
	\title{EVT-enriched Radio Maps for URLLC
}
	\author{
        \IEEEauthorblockN{Dian Echevarría Pérez, \IEEEmembership{Graduate Student Member, IEEE} \IEEEmembership{} %
        Onel L. Alcaraz López, \IEEEmembership{Member, IEEE}, Hirley Alves, \IEEEmembership{Member, IEEE}
        } 
        
		\thanks{The authors are with the Centre for Wireless Communications (CWC), University of Oulu, Finland. \{dian.echevarriaperez,  onel.alcarazlopez, hirley.alves\}@oulu.fi}

        \thanks{This research was supported by the Research Council of Finland (former Academy of Finland) 6G Flagship Programme (Grant Number: 346208), and the Finnish Foundation for Technology Promotion.}
    }  
    \maketitle
	\begin{abstract}
  This paper introduces a sophisticated and adaptable framework combining extreme value theory with radio maps to spatially model extreme channel conditions accurately. Utilising existing signal-to-noise ratio (SNR) measurements and leveraging Gaussian processes, our approach predicts the tail of the SNR distribution, which entails estimating the parameters of a generalised Pareto distribution, at unobserved locations. This innovative method offers a versatile solution adaptable to various resource allocation challenges in ultra-reliable low-latency communications. We evaluate the performance of this method in a rate maximisation problem with defined outage constraints and compare it with a benchmark in the literature. Notably, the proposed approach meets the outage demands in a larger percentage of the coverage area and reaches higher transmission rates.
\end{abstract}
\begin{IEEEkeywords}
    availability, extreme value theory, radio maps, reliability,  URLLC.
\end{IEEEkeywords}

\section{Introduction}
Ultra-reliable low-latency communication (URLLC) is an essential operation mode for current and future wireless communications networks. Simultaneously achieving high reliability, \textit{e.g.,} $10^{-9}-10^{-3}$ error rates, and low latency, \textit{e.g.,} 1~ms, is extremely challenging. It requires efficient resource allocation mechanisms leveraging knowledge about the distribution tail of the signal-to-noise ratio (SNR) and the spatial characteristics of the environment \cite{Mahmood.2020,lopez2022statistical}. Interestingly, characterising the distribution tail and occurrence of extreme events is the extreme value theory (EVT)'s main focus. EVT provides statistical modelling for phenomena critical to URLLC, phenomena that central limit theorem-based approaches fail to capture accurately \cite{coles2001introduction}. On the other hand, spatial features of the environment, such as spatial correlation and probability of line-of-sight (LOS), are efficiently modelled using radio maps. Indeed, radio maps are powerful tools for representing various characteristics of the radio environment over a geographical area\cite{RAdioMaps1}.   
 
\subsection{EVT for URLLC}
In wireless communications, EVT models and predicts extreme conditions that can significantly impact the system's reliability. For instance, the work in \cite{perez2023extreme} presented an EVT-based algorithm to solve a minimum-power precoding problem with outage constraints in the presence of imperfect channel state information (CSI). The authors also analysed the impact of the number of samples on the system performance and showed that with fewer samples than $1/\epsilon$, the target outage probability $\epsilon$ could be met. The authors in \cite{mehrnia2021wireless} presented a methodology for URLLC that fits the lower tail distribution of the received power in a wireless channel and determined the optimal threshold value of the Generalized Pareto Distribution (GPD). Moreover, they showed that their proposed approach can considerably reduce the number of samples required to fit the data. In \cite{samarakoon2019distributed}, the authors proposed a scheme that combines federated learning and EVT to learn the network statistics and exploit them to reduce the occurrence of large queues while minimising the transmit power. An EVT-rate selection approach for URLLC was presented in \cite{mehrnia2022extreme}. The authors fitted the distribution tail of the received power to the GPD and determined the maximum rate that guarantees the outage requirements. Finally, the work in \cite{EVTMehrnia} proposed an EVT-based channel modelling methodology for estimating the multivariate channel tail statistics of a multiple-input, multiple-output URLLC system. They showed that the proposed multi-dimensional channel modelling approach better fits the empirical data in the lower tail than conventional extrapolation-based approaches.

\subsection{Radio Maps and applications for URLLC}

The radio maps tool directly applies to wireless communications by providing detailed representations of signal characteristics across various geographical areas. Indeed, it offers unparalleled insights into network behaviour and performance and paves the way for optimising wireless network design and deployment, ensuring enhanced connectivity and service quality. Several works have recently proposed different methods to construct and exploit radio maps for various resource allocation problems. For instance, the work in \cite{8344806} presented a distributed algorithm based on regression Kriging for radio map reconstruction regarding average received power at locations without sensor measurements. The algorithm minimises the number of sensor measurements required
for radio map reconstruction through distributed processing and clustering of sensor nodes. The authors in \cite{radiomapWang} proposed a deep Gaussian process for indoor radio map construction and location estimation. They used samples of received signal strength to generate high-resolution radio maps at unobserved locations. The work in \cite{9634111} focused on constructing radio maps for cellular systems with massive directional antenna arrays. They proposed a technique based on semi-parametric Gaussian regression that outperforms parametric and non-parametric radio map construction for map generation. In the context of URLLC, the authors in \cite{10105152} compared radio maps and channel charting performance to predict channel capacity at unobserved locations. They showed that the radio-map-based approach outperforms channel charting not only in predicted channel capacity but also in terms of outage probability. Finally, the work in \cite{kallehauge2022predictive} exploited Gaussian processes for the radio map generation of the $\epsilon-$quantile of the logarithmic SNR at unobserved locations. They exploited the radio map to solve a rate selection problem with defined outage constraints. The proposed method was compared with a baseline scheme that exploits the quantile of the nearest observation for the rate selection, showing the superiority of the proposed algorithm.

\subsection{Contributions}

While radio maps provide a broad spatial analysis, they often fail to accurately model the tail distribution of the SNR, crucial for the stringent resource allocation demands of URLLC. In contrast, EVT excels in predicting rare events at locations with rich data but struggles in areas without extensive measurements. Therefore, utilising EVT to enhance the characterisation of extreme values, combined with the predictive spatial insights from radio maps, emerges as a strategic approach to strengthen the network everywhere. This integration promises to create a thorough and resilient framework for URLLC, ensuring robust performance across all locations.

The main contributions of this work are summarised as follows:
\begin{itemize}
    \item We introduce a novel framework that integrates EVT with radio maps to spatially model extreme channel conditions accurately. Our methodology exploits existing SNR measurements and Gaussian processes to estimate the SNR distribution tail via GPD parameters at unobserved locations within the coverage area. 
    \item We demonstrate the versatility of our proposed method by applying it to a rate maximisation problem with outage constraints. We highlight its adaptability to solve multiple resource allocation problems within the URLLC framework.
    \item Through comprehensive simulations, we compare our method with the approach presented in \cite{kallehauge2022predictive} which exploits SNR quantile predictions for the rate selection, demonstrating that our proposed method meets outage demands across a larger percentage of the coverage area and achieves higher transmission rates.
    \item We show that utilising EVT reduces the number of required samples for accurate prediction and optimisation compared to the benchmark and other existing approaches. This efficiency in sample usage further underscores the practical applicability and efficiency of our proposed framework.
\end{itemize}
\textbf{Notation} Superscript $(\cdot)^T$ denotes the transpose operator, $(\cdot)^{-1}$ represents the matrix inverse operation, and $||\cdot||$ depicts the norm of a vector. Moreover, $\lfloor\cdot\rfloor$ represents the floor operator and $\mathcal{N}(\mathbf{v},\mathbf{R})$ denotes a Gaussian distribution with mean vector $\mathbf{v}$ and covariance matrix $\mathbf{R}$. $Q^{-1}(\cdot)$ and $\text{erf}^{-1} (\cdot)$ represent the inverse Q-function and inverse error function, respectively. Finally, $\mathcal{Q}(c,D)$ depicts the $c\%$-quantile operator of the sample set $D$.  Table \ref{table_0} summarises the main symbols used throughout the paper.

\section{System model}\label{Sect_System}

\begin{figure}[t!]
    \centering  \includegraphics[width=\columnwidth]{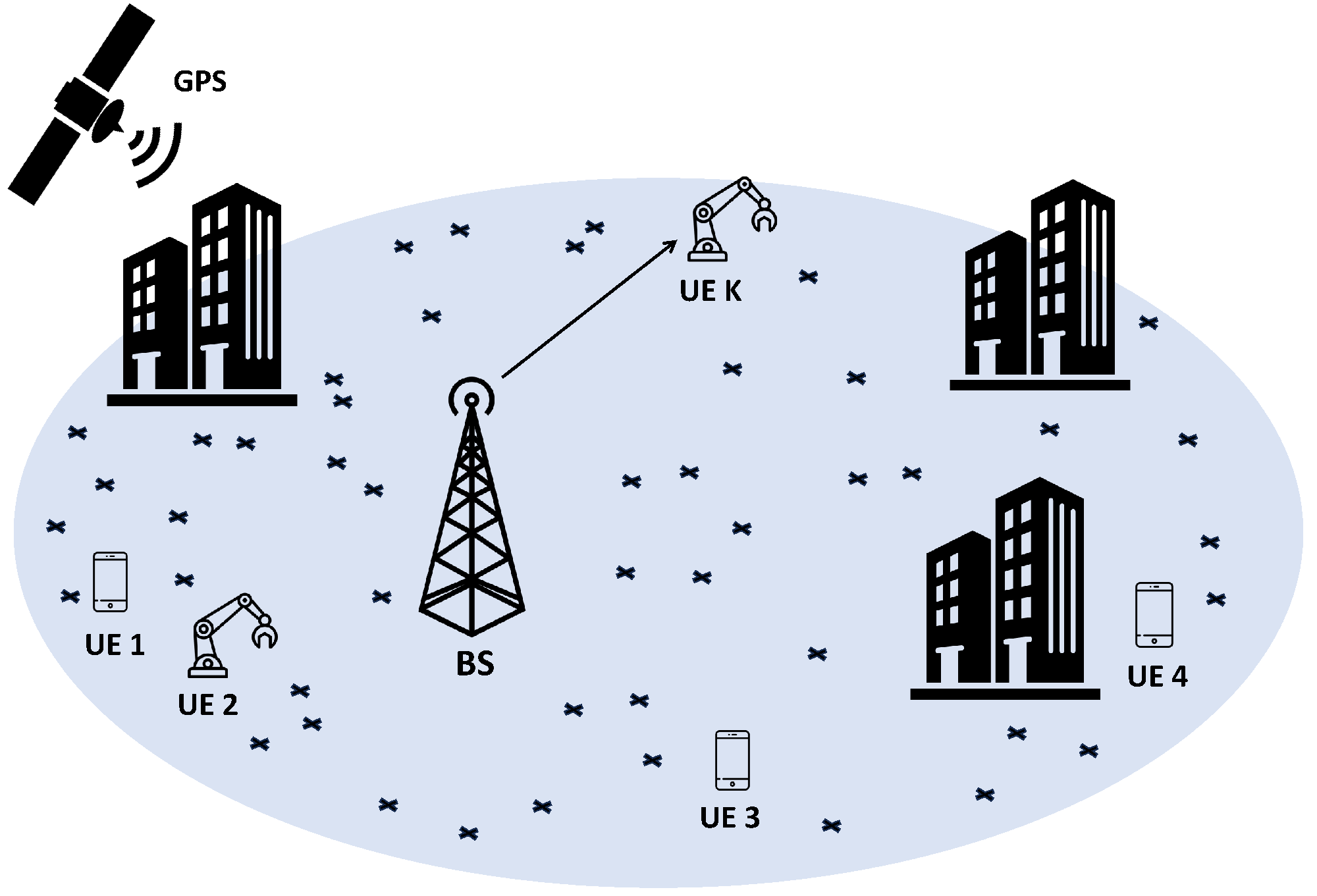}
    \caption{System model. The BS serves $K$ UEs in the DL, guaranteeing URLLC quality-of-service demands. The crosses depict the locations where SNR measurements are available from previous UEs in the network. Notice that the density of observation points in the figure does not match the number of localisations with available samples in a real scenario and is only used for visualisation purposes. The GPS provides the localisation of the SNR measurements and UEs.}
    \label{syst_model}
\end{figure}

We consider a single-antenna base station (BS) that serves $K$ single-antenna URLLC user equipment (UE) in the downlink (DL) within its coverage area (see Fig.~\ref{syst_model}). The BS knows a history of $N$ independent and identically distributed (i.i.d.) SNR measurements at $M$ different locations acquired from previous transmissions, \textit{i.e.,} $\Upsilon(l_m) = \{\gamma_{1}(l_m), \gamma_{2}(l_m),...,\gamma_{N}(l_m)\}$ where $l_m = [x_m, y_m] \ \forall m~\in~[1,M]$ represents the geographic coordinates of the measurements. The i.i.d. assumption can be justified by setting the sampling period larger than the coherence time. The BS perfectly knows the coordinates of the measurements, \textit{i.e.,} $\mathcal{L} = [l_1 \ l_2...\ l_M]^T$ as well as the coordinates of any UE $k$, \textit{i.e.,} $l_k = [x_k, y_k]$. Such coordinates may be acquired using, for instance, the global positioning system (GPS). When the BS is transmitting, the perceived SNR at location $l_m$ is given by
\begin{equation}\label{SNR}
   \gamma(l_m) = \frac{p|h(l_m)|^2}{\upsilon^2},
\end{equation}
where $p$ represents the transmit power, $h(l_m), $ depicts the channel coefficient between the BS and the antenna of a UE at location $l_m$, which captures the effect of both large-scale and small-scale fading, and $\upsilon^2$ represents the noise variance.

\begin{table}[t!]
    \centering
    \caption{Main symbols used throughout the paper}
\label{table_0}    \begin{tabular}{l l}
        \hline
        \textbf{Symbol} & \textbf{Definition} \\
        \hline
        $M$ & Number of locations with SNR measurements \\
          $\Upsilon$ & Set the SNR measurements\\ 
          $R_m$ & Transmission rate\\
        $N$ & Number of independent i.i.d. SNR measurements \\
        $\gamma$ & Achieved SNR\\
        $p$ & Transmit power \\
        $\upsilon^2$ & Noise variance \\
        $h$ & Complex channel coefficient\\
        $O$ & Outage probability\\
        $\gamma^{\text{tar}}$ & SNR target \\
        $\zeta$ & Target outage probability \\
        $\mu$, $\xi$, $\sigma$ & Threshold, shape and scale parameters of the GPD \\
        $\rho$ & quantile value of the samples\\
        \hline
    \end{tabular}
\end{table}
\subsection{Problem definition}




Our main goal is to propose a general framework that can be used to find solutions for optimisation problems with link-reliability constraints. The reliability is commonly measured in terms of the outage probability $\mathcal{O}$,   \textit{i.e.,}
\begin{equation}\label{outage_eq}
    \mathcal{O}(l_m)=\text{Pr}\big\{\gamma(l_m)<\gamma^{tar}(l_m)\big\}\le\zeta,
\end{equation}
 where $\gamma^{tar}(l_m)$ depicts the required SNR to achieve a successful transmission at location $l_m$, and $\zeta\ll1$ represents the target outage probability. Specifically, we exploit the proposed method in \cite{kallehauge2022predictive} for radio map construction using Gaussian processes and tune it according to our specific model. We also exploit the EVT-based approach in \cite{lopez2022statistical, perez2023extreme} to deal with extreme events. We integrate both approaches into a framework for URLLC based on EVT and radio maps.
 \section{EVT and radio maps integration}\label{EVT-Radio}

 \subsection{EVT preliminaries}
 The main outcome we exploit from EVT is the following: 

\begin{theorem}[Theorem for Exceedances Over Thresholds \cite{coles2001introduction}]
\label{th:exceedance-th-coles}
For a given random variable $X$ from a non-degenerative distribution and for a large enough threshold $\mu$, the cumulative distribution function (CDF) of $Z =X-\mu$ conditioned on $X>\mu$ is given by
\begin{align}\label{GPD}
    F_Z(z) = 1-\bigg[1+\frac{\xi z}{\sigma}\bigg]^{-\frac{1}{\xi}},
\end{align}
defined on $\{z: z>0 \ \text{and} \ 1+\xi z/\sigma > 0\}$. The distribution in \eqref{GPD} is known as the \textit{GPD} with shape and scale parameters $\xi$ and  $\sigma$, respectively.
\end{theorem}

Multiple random variable $X$ observations are required to estimate the GPD parameters effectively. Log-likelihood methods are common for estimating $\xi$ and $\sigma$. On the other hand, the parameter $\mu$ can be obtained from the \textit{mean residual life method} or from a complementary approach termed as \textit{parameter stability method} \cite{coles2001introduction}. Also, the \textit{fixed threshold approach}, with the threshold typically defined before fitting, is commonly used. The upper 10\% rule of DuMouchel is a practical example, which uses up to the upper 10$\%$ of the data to fit the GPD, \textit{i.e.,} $\mu = \mathcal{Q}(\rho\times 100,X)$ with $\rho\ge0.9$\cite{dumouchel1983estimating}.
 \subsection{EVT-based constraint reformulation}\label{evt-reformulation}
 Notice that the SNR distribution is unknown in practical systems. Fortunately, knowing the underlying distribution of the SNR data is not required to apply EVT effectively. Thus, using the principle in \textbf{Theorem 1} to model the SNR distribution tail and reformulate \eqref{outage_eq} seems appealing. Let us proceed as follows at each location $l_m$ \cite{perez2023extreme}
\begin{align}\label{step1}
    \text{Pr}\big\{\gamma({l_m})\!<\!\gamma^{tar}({l_m})\big\}\!=\!\text{Pr}\big\{f\big(\gamma({l_m})\big)\!>\!f(\gamma^{tar}(l_m)\big)\big\},
\end{align}
which comes from applying a concave function $f(\cdot)$ that handles dispersed data and performs a mirroring operation since we need the data to be in the right tail to apply the principle in \textbf{Theorem 1}. Exploiting the SNR measurements 
 $\Upsilon(l_m)$ and for ease of notation,  we define $\psi_{l_m} \triangleq f\big(\Upsilon({l_m})\big)$ and $\phi_{l_m} \triangleq f\big(\gamma^{tar}({l_m})\big)$. Defining a threshold $\mu(l_m)$ and applying the principle of conditional expectation, we have
\begin{align}\label{step2}
     &\text{Pr}\big\{\psi_{l_m}>\phi_{l_m}\big\} \nonumber\\ &  = \text{Pr}\big\{\!\psi_{l_m}\!\!>\!\!\mu(l_m)\!\big\}\text{Pr}\big\{\!\psi_{l_m}\!\!-\!\mu(l_m) \!\!>\!\!\phi_{l_m}\!\!-\!\mu(l_m) \big| \psi_{l_m}\!>\!\mu(l_m)\!\big\}.
\end{align}
Estimating the threshold $\hat{{\mu}}(l_m)$ according to the DuMouchel's rule, we have 
\begin{equation}\label{mu_def}
\hat{\mu}(l_m)=\mathcal{Q}\big(\rho\times 100, \psi_{l_m}\big),
\end{equation}
therefore, 
\begin{equation}\label{prob1}
    \text{Pr}\big\{\!\psi_{l_m}>\hat{\mu}(l_m)\!\big\} = 1-\rho.
\end{equation}
Computing the excess data $\psi_{l_m}-\hat{\mu}(l_m)\big| \psi_{l_m}\!>\!\hat{\mu}(l_m)$,
 and obtaining log-likelihood estimates of shape, $\hat{\xi}(l_m)$, and scale, $\hat{\sigma}(l_m)$, \eqref{outage_eq} can be rewritten as 
\begin{equation}\label{Outage_eq_2}
    \mathcal{O}(l_m) = (1-\rho)\Big(1+\frac{\hat{\xi}(l_m)}{\hat{\sigma}(l_m)}\big(\phi_{l_m} -\hat{\mu}(l_m)\big)\Big)^{-1/\hat{\xi}(l_m)}.
\end{equation}
The accuracy of this equation increases with the number of samples $N$, meaning that the estimates converge to the actual parameters as $N\rightarrow\infty$ \cite{rice2006mathematical}.
\subsection{Radio map construction} \label{radio-gen}
The above equation provides a comprehensive characterisation of the tail region of the SNR at the observed locations $\mathcal{L}$. However, the number of locations without SNR observation $\mathcal{L'} = [l_1', l_2',...\ l_M']$ is typically larger and almost impossible to cover in practice with measurements. In this sense, predicting the parameters $\{\hat{\xi}(l_m'), \hat{\sigma}(l_m'),\hat{\mu}(l_m')\}$ at unknown locations based on the observed data and exploiting the spatial correlation of the environment seems more appealing. 

The work in \cite{kallehauge2022predictive} presented a Gaussian-process-based scheme for constructing a radio map of quantiles of the logarithmic SNR from observed measurements exploiting the spatial correlation of the SNR. Surprisingly, a quantile is the principle behind DuMouchel's rule and, therefore, the way to determine the threshold values $\hat{\mu}$ in \eqref{Outage_eq_2}. The distribution of the quantiles is asymptotically Gaussian \cite{stuart2010kendall} while scale and shape parameters are also Gaussian-distributed when log-likelihood estimation methods are used \cite{rice2006mathematical}. Therefore, it is expected that maps of $\hat{\xi}$ and $\hat{\sigma}$ can be constructed using Gaussian processes. Notice that these parameters characterise the tail region of the SNR. Thus, their observations will also present a significant correlation across space as with the quantiles.

Let us create a map for each parameter $\hat{\xi}$, $\hat{\sigma}$ and $\hat{\mu}$ following the procedure in \cite{kallehauge2022predictive}. For each dataset $\Upsilon(l_m)$, we compute the quantile $\hat{\mu}(l_m)$ of $f(\Upsilon(l_m))$, obtain the excess data, and get log-likelihood estimates $\hat{\xi}(l_m)$ and $\hat{\sigma}(l_m)$ as previously discussed. We now define three new sets in vector form as 
\begin{align}  
\hat{\mathbf{e}}(\mathcal{L}) &= [\hat{\xi}(l_1)\  \hat{\xi}(l_2)  \ ...\ \hat{\xi}(l_M)]^T,\nonumber\\ \hat{\mathbf{s}}(\mathcal{L}) &= [\hat{\sigma}(l_1)  \ \hat{\sigma}(l_2)\ ...\ \hat{\sigma}(l_M)]^T, \nonumber \\ \hat{\mathbf{u}}(\mathcal{L}) &= [\hat{\mu}(l_1) \ \hat{\mu}(l_2)\ ...\ \hat{\mu}(l_M)]^T. \nonumber 
\end{align}

Next, compute the sample mean and sample standard deviation of each parameter as follows
\begin{align}
    \label{mean}
    \bar{\mathcal{J}} &= \frac{1}{M}\sum_{m = 1}^{M}  \hat{\mathcal{J}}(l_m),\\
\label{std}
 {\bar{\bar{\mathcal{J}}}} &= \Bigg[\frac{1}{M-1}\sum_{m = 1}^{M} \Big(\hat{\mathcal{J}}(l_m) -\bar{\mathcal{J}}\Big)^2\Bigg]^{\frac{1}{2}},
\end{align}
where $\bar{\mathcal{J}} \rightarrow \big\{\bar{\xi}, \bar{\sigma}, \bar{\mu}\big\}$, $\bar{\bar{\mathcal{J}}} \rightarrow \big\{\bar{\bar{\xi}}, \bar{\bar{\sigma}}, \bar{\bar{\mu}}\big\}$ and $\hat{\mathcal{J}}(l_m) \rightarrow \big\{\hat{\xi}(l_m), \hat{\sigma}(l_m), \hat{\mu}(l_m)\big\}$. 
The sets  $\hat{\mathbf{e}}(\mathcal{L})$, $\hat{\mathbf{s}}(\mathcal{L})$ and $\hat{\mathbf{u}}(\mathcal{L})$ must be normalized such that each entry $\mathcal{J}^{\circ}(l_m) \rightarrow \big\{\xi^{\circ}(l_m),\sigma^{\circ}(l_m), \mu^{\circ}\!(l_m)\big\}$ is given by
\begin{equation}\label{normalization}
 \mathcal{J}^{\circ}(l_m)= \frac{\hat{\mathcal{J}}(l_m)-\bar{\mathcal{J}}}{{\bar{\bar{\mathcal{J}}}}},
\end{equation}
which allows to form new sets as 
\begin{align}
\mathbf{e}^{\circ}(\mathcal{L}) =& [\xi^{\circ}(l_1)\  \xi^{\circ}(l_2)  \ ...\ \xi^{\circ}(l_M)]^T,\nonumber\\ \mathbf{s}^{\circ}(\mathcal{L}) =& [\sigma^{\circ}(l_1)\  \sigma^{\circ}(l_2)  \ ...\ \sigma^{\circ}(l_M)]^T,\nonumber\\ \mathbf{u}^{\circ}(\mathcal{L}) =& [\mu^{\circ}(l_1)\  \mu^{\circ}(l_2)  \ ...\ \mu^{\circ}(l_M)]^T. \nonumber
\end{align}
Assuming the observation model in \cite{kallehauge2022predictive}, we have
\begin{align}\label{models}
     \mathcal{J}^{\circ}(l_m) & = \mathcal{J}(l_m)+\Tilde{\mathcal{J}},
\end{align}
where $\mathcal{J}(l_m) \rightarrow \big\{\xi(l_m),\sigma(l_m),\mu(l_m)\big\}$ and  $\Tilde{\mathcal{J}}\sim\mathcal{N}\big(0,\lambda^2_{\Tilde{\mathcal{J}}}\big)$ with $\Tilde{\mathcal{J}} \rightarrow \big\{\Tilde{\xi}, \Tilde{\sigma}, \Tilde{\mu}\big\}$ representing the zero-mean independent observation noise with variance $\lambda^2_{\Tilde{\mathcal{J}}}$. Moreover, the variables $\xi$, $ \sigma$, and $\mu$ represent Gaussian processes, thus, 
\begin{align}
    \mathbf{e}(\mathcal{L}) &= [\xi(l_1)\  \xi(l_2)  \ ...\ \xi(l_M)]^T,\nonumber \\ 
    \mathbf{s}(\mathcal{L}) &= [\sigma(l_1)  \ \sigma(l_2)\ ...\ \sigma(l_M)]^T, \nonumber\\
    \mathbf{u}(\mathcal{L}) &= [\mu(l_1) \ \mu(l_2)\ ...\ \mu(l_M)]^T, \nonumber
\end{align}

    are jointly Gaussian vectors such that \cite{williams2006gaussian}
\begin{align}\label{GP}
\mathbf{d}(\mathcal{L})&\sim\mathcal{N}(\mathbf{0}, \mathbf{C}^{d}_{\scaleto{\mathcal{L}\mathcal{L}}{4pt}}),
\end{align}
where $\mathbf{d}(\mathcal{L}) \rightarrow \{\mathbf{e}(\mathcal{L}),\mathbf{s}(\mathcal{L}),\mathbf{u}(\mathcal{L})\}$. The matrix $ \mathbf{C}^{d}_{\scaleto{\mathcal{L}\mathcal{L}}{4pt}}$ captures the spatial correlation between observations at the different positions 
$\mathcal{L}$. The Gudmundson correlation model is commonly used for modelling the variations of the SNR in the space\cite{gudmundson1991correlation}, thus, suitable for modelling $ \mathbf{C}^{\scaleto{\mathcal{U}}{4pt}}_{\scaleto{\mathcal{L}\mathcal{L}}{4pt}}$. The entries of the matrix are given by
\begin{equation}\label{Gudmundson}
\mathbf{C}^{u}_{\scaleto{\mathcal{L}\mathcal{L}}{4pt}}(i,j) = \omega_{u}^2\exp\Big(\frac{||l_i-l_j||}{r_{u}}\Big) \ \ \ i,j \in [1, N],
\end{equation}
where $\omega_{u}^2$ and $r_{u}$ represent the variance of the process and decorrelation distance, respectively. Another commonly used model to capture spatial correlation is the Matérn model, which is particularly useful due to an additional parameter that controls the smoothness of the correlation function \cite{williams2006gaussian}. Interestingly, we found through numerical experimentation that this model captures the spatial behaviour of $\mathbf{e}(\mathcal{L})$ and $\mathbf{s}(\mathcal{L})$ more accurately than Gudmundson's model. In this case, the matrix structures are given by

\begin{align}
\mathbf{C}^{e}_{\scaleto{\mathcal{L}\mathcal{L}}{4pt}}(i,j) \! = \!\omega_{e}^2\frac{2^{1\!-\!\nu_{e}}}{\Gamma({\nu_{e}})}\!\Bigg(\!\!\frac{\sqrt{\nu_{e}}||l_i\!-\!l_j||}{r_{e}}\!\Bigg)^{\nu_{e}}\!\!\!\!\!&B_{\nu_{e}}\!\Bigg(\!\!\frac{\sqrt{\nu_{e}}||l_i\!-\!l_j||}{r_{e}}\!\Bigg), \label{Covariance1}\\
\mathbf{C}^{s}_{\scaleto{\mathcal{L}\mathcal{L}}{4pt}}(i,j) \! = \!\omega_{s}^2\frac{2^{1\!-\!\nu_{s}}}{\Gamma({\nu_{s}})}\!\Bigg(\!\!\frac{\sqrt{\nu_{s}}||l_i\!-\!l_j||}{r_{s}}\!\Bigg)^{\nu_{s}}\!\!\!\!\!&B_{\nu_{s}}\!\Bigg(\!\!\frac{\sqrt{\nu_{s}}||l_i\!-\!l_j||}{r_{s}}\!\Bigg)\label{Covariance2},
\end{align}
where $i, j\in[1, N]$,  \{$\omega_{e}^2,\omega_{s}^2\}$ represent the variances of the processes, $\{\nu_{e},\nu_{s}\}$ control the smoothness of the functions, $\{r_{e},r_{s}\}$ capture the correlation decays with the distance, and $\Gamma(\cdot)$ and $\big\{B_{\nu_{e}}(\cdot),B_{\nu_{s}}(\cdot)\big\}$ depict the Gamma function and modified Bessel function of second order, respectively. To construct a map at a given set of $M'$ unobserved locations $\mathcal{L'} = [l_1' \ l_2'...\ l_{M'}']^T$ (regular grid), and assuming that the joint distribution of normalised observations $\mathbf{d}^\circ(\mathcal{L}) \rightarrow \{\mathbf{e}^\circ(\mathcal{L}),\mathbf{s}^\circ(\mathcal{L}),\mathbf{u}^\circ(\mathcal{L})\}$ and predictions $\mathbf{d}^\circ(\mathcal{L'}) \rightarrow \{\mathbf{e}^\circ(\mathcal{L}'),\mathbf{s}^\circ(\mathcal{L}'),\mathbf{u}^{\circ}(\mathcal{L}')\}$ is multivariate Gaussian, we have \begin{equation}
\begin{bmatrix}
\mathbf{d}^\circ(\mathcal{L'}) \\
\mathbf{d}^\circ(\mathcal{L})
\end{bmatrix}
\sim \mathcal{N} \left( \mathbf{0}, \begin{bmatrix}
 \mathbf{C}^{d}_{\scaleto{\mathcal{L'}\mathcal{L'}}{4pt}} &  \mathbf{C}^{d}_{\scaleto{\mathcal{L'}\mathcal{L}}{4pt}} \\
 \mathbf{C}^{d}_{\scaleto{\mathcal{L}\mathcal{L'}}{4pt}} &  \mathbf{C}^{d}_{\scaleto{\mathcal{L}\mathcal{L}}{4pt}} + \lambda^2_{\Tilde{\mathcal{J}}}\mathbf{I}_{M},
\end{bmatrix} \right).
\end{equation}
The conditional mean and covariance of the predicted variables are given by \cite{kallehauge2022predictive, williams2006gaussian}
\begin{align}
\mathbf{m}^{d(\mathcal{L}')} &= \mathbf{C}^{d}_{\scaleto{\mathcal{L'}\mathcal{L}}{4pt}} (\mathbf{C}^{d}_{\scaleto{\mathcal{L}\mathcal{L}}{4pt}} + \omega^2_{d} \mathbf{I}_M)^{-1}\mathbf{d}^{\circ}(\mathcal{L}),\label{meanV}\\
\mathbf{C}^{d(\mathcal{L}')}& = \mathbf{C}^{d}_{\scaleto{\mathcal{L'}\mathcal{L'}}{4pt}} - \mathbf{C}^{d}_{\scaleto{\mathcal{L'}\mathcal{L}}{4pt}} (\mathbf{C}^{d}_{\scaleto{\mathcal{L}\mathcal{L}}{4pt}} + \lambda^2_{\Tilde{\mathcal{J}}}\mathbf{I}_{M})^{-1} \mathbf{C}^{d}_{\scaleto{\mathcal{L}\mathcal{L}}{4pt}}\label{meanC}.
\end{align}
Since \eqref{meanV} and \eqref{meanC} are obtained for normalised data, we must proceed to denormalise as follows 
\begin{align}
    \mathbf{m}^{\hat{d}(\mathcal{L}')}& = \mathbf{m}^{d(\mathcal{L}')}\bar{\bar{\mathcal{J}}}+\bar{\mathcal{J}},\label{mean2}\\
    \mathbf{C}^{\hat{d}(\mathcal{L}')}&=\mathbf{C}^{d(\mathcal{L}')}\bar{\bar{\mathcal{J}}}^2.\label{cov2}
\end{align}
The expressions in \eqref{mean2} and \eqref{cov2} provide a statistical characterisation of the parameters of the GPD at unobserved locations whose accuracy we test in Section \ref{section_4}. Notice that the parameters $\omega_d^2$, $r_d$, $\nu_d$ and $\lambda_{\mathcal{\Tilde{J}}}^2$ must be estimated in practice from the available data, for instance, with log-likelihood estimation.  
\subsection{Definition of predictive constraint}\label{predictiveConst}
We obtain enough information to define the predictive outage probability expressions at unobserved locations with the steps followed in Sections \ref{evt-reformulation} and \ref{radio-gen}. Let us define the predicted variables as 
\begin{align}
    \hat{\mathbf{e}}(\mathcal{L}') &=  \mathbf{m}^{\hat{e}(\mathcal{L}')} = [\hat{\xi}(l_1')\  \hat{\xi}(l_2')  \ ...\ \hat{\xi}(l_M')]^T,\label{e1}\\ \hat{\mathbf{s}}(\mathcal{L}') &=  \mathbf{m}^{\hat{s}(\mathcal{L}')} =[\hat{\sigma}(l_1')  \ \hat{\sigma}(l_2')\ ...\ \hat{\sigma}(l_M')]^T,\label{s1}\\
     \hat{\mathbf{u}}(\mathcal{L}') &=  \mathbf{m}^{\hat{u}(\mathcal{L}')} =[\hat{\mu}(l_1') \ \hat{\mu}(l_2')\ ...\ \hat{\mu}(l_M')]^T\label{u1}.
\end{align}
Notice that the predictions can vary around the mean within certain margin values defined by the variances contained in the diagonal elements of the matrices $\mathbf{C}^{\hat{d}(\mathcal{L}')}$. Computing the confidence margins as $\tau-$quantiles ensures robustness for the predictions. For each entry $m'$ in
$\hat{\mathbf{u}}(\mathcal{L}')$, we have 
\begin{equation}\label{quantile} 
 \hat{\mu}^{\tau}(l_m') = Q^{-1}\Big(\tau, \hat{\mathbf{u}}_{m'}(\mathcal{L}'), \mathbf{C}^{\hat{u}(\mathcal{L}')}(m',m')\Big),   
\end{equation}
where $Q^{-1}(\cdot)$ represents the inverse Q-function. The variances of $\hat{\xi}(l_m)$ and $\hat{\sigma}(l_m)$ are expected to be significantly smaller than that of $\hat{\mu}(l_m)$ due to their typical value ranges, thus, inserting margins
for them will cause little/no variations in the outline of the distribution tail as shown through simulations in Section \ref{section_4}. Therefore, the predictive outage probability expression at each unobserved location $l_m'$ is given by

\begin{equation}\label{Outage_eq_3}
    \mathcal{O}(l_m') = (1-\rho)\Big(1+\frac{\hat{\xi}(l_m')}{\hat{\sigma}(l_m')}\big(\phi_{l_m'} -\hat{\mu}^{\tau}(l_m')\big)\Big)^{-1/\hat{\xi}(l_m')}.
\end{equation}
Notice that the parameter $\phi_{l_m'}$ allows to control the rate and/or transmit power, thus essential for rate/power allocation problems with outage constraints.

\section{Optimization framework}
To evaluate the performance of the proposed framework, we consider the following optimisation problem at every location  $l_{m''} \in \mathcal{L}''=\mathcal{L}' \cup \mathcal{L}$ in the coverage area
\begin{subequations}\label{P1}
	\begin{alignat}{2}
	\mathbf{P1:}\qquad &\underset{\gamma^{tar}(l_{m''})}{\mathrm{maximise}}  &\ \ \ &\
	R_m(l''  ) \label{P1:a}\\
	&\text{subject to}   &      & \mathcal{O}(l_m'') \leq \zeta, \ \ \forall m'', \label{P1:b}
	\end{alignat}
\end{subequations}
where $R_m = \log_2(1+\gamma(l_m''))$ represents the spectral efficiency in bps/Hz. The problem aims to maximise the spectral efficiency (termed as rate hereinafter) in any location $l_{m''}$ where a UE $k$ might be located $(l_{m''} = l_k)$ while guaranteeing strict outage requirements. 

\subsection{Proposed algorithm}
For solving $\mathbf{P1}$, we proceed as described in Sections \ref{evt-reformulation}, \ref{radio-gen} and \ref{predictiveConst} to obtain \eqref{Outage_eq_3} for every location $l_{m''}$ in the map. The goal is to find the maximum target SNR $\gamma^{tar}(l_{m''})$ that guarantees the target outage probability $\zeta$. Thus, substituting $\mathcal{O}(l_{m''})$ by $\zeta$ in \eqref{Outage_eq_3} and isolating $\phi_{l_{m''}}$, we have
\begin{equation}\label{PsiEq}
    \phi_{l_{m''}} = \frac{\hat{\sigma}(l_{m''})}{\hat{\xi}(l_{m''})}\Bigg[\bigg[\frac{\zeta}{1-\rho}\bigg]^{-\hat{\xi}(l_{m''})}-1\Bigg]+\hat{\mu}^{\tau}(l_{m''}),
\end{equation}
then, the rate is given as
\begin{equation}\label{rate_eq}
    R_m(l_{m''}) = \log_2\Big(1+f^{-1}\big(\phi_{l_{m''}}\big)\Big),
\end{equation}
All the discussed steps for the rate selection problem are summarised in \textbf{Algortihm 1}. Notice that a similar approach can be followed for other problems, such as transmit power minimisation with outage constraints.

\begin{algorithm}[t!]
\caption{Rate maximisation for URLLC}
\hspace*{\algorithmicindent} \textbf{Inputs:} $\rho,\ \{\Upsilon(l_m)\}, \  \mathcal{L},\ \mathcal{L{''}}$, $\zeta$, $\tau$ \\
\hspace*{\algorithmicindent}
\textbf{Outputs:} $\ \{R_m\}$
\begin{algorithmic}[1]
\State Compute $\psi_{l_m} =f\big(\Upsilon({l_m})\big)$ and $ \hat{\mu}(l_m)$ with \eqref{threshold} $\forall l_m\in\mathcal{L}$ 
\State Find the excess data $\psi_{l_m}-\hat{\mu}(l_m)\big| \psi_{l_m}\!>\!\hat{\mu}(l_m) \ \forall l_m\in\mathcal{L}$
\State Define the sets $\hat{\mathbf{e}}(\mathcal{L}), \hat{\mathbf{s}}(\mathcal{L}) $ and $ \hat{\mathbf{u}}(\mathcal{L})$
\State Obtain normalized sets $\hat{\mathbf{e}}^\circ(\mathcal{L}), \hat{\mathbf{s}}^\circ(\mathcal{L}) $ and $ \hat{\mathbf{u}}^\circ(\mathcal{L})$ with \eqref{mean}, \eqref{std} and \eqref{normalization}
\State Obtain log-likelihood estimates of  $\omega_d^2$, $r_d$, $\nu_d$ and $\lambda_{\mathcal{\Tilde{J}}}^2$ 
\State Compute the covariance matrices of $\mathbf{u}(\mathcal{L})$, $\mathbf{e}(\mathcal{L})$ and $\mathbf{s}(\mathcal{L})$ with \eqref{Gudmundson}, \eqref{Covariance1} and \eqref{Covariance2}, repectively
\State Compute predictive mean and covariance of $\hat{\mathbf{u}}(\mathcal{L''})$, $\hat{\mathbf{e}}(\mathcal{L''})$ and $\hat{\mathbf{s}}(\mathcal{L''})$ with \eqref{meanV} and \eqref{meanC}
\State Denormalize predictive mean and covariance of $\hat{\mathbf{u}}(\mathcal{L''})$, $\hat{\mathbf{e}}(\mathcal{L''})$ and $\hat{\mathbf{s}}(\mathcal{L''})$ with \eqref{mean2} and \eqref{cov2}
\State Obtain $\hat{\xi}(l_{m''})$ and $\hat{\sigma}(l_{m''})$ $\forall l_{m''}\in\mathcal{L}''$ as the entries of \eqref{e1} and \eqref{s1}, respectively
\State For each entry of $\hat{\mathbf{u}}(\mathcal{L''})$ in \eqref{u1} compute the $\tau-$quantile $\hat{\mu}^{\tau}(l_{m''})$  with \eqref{quantile}
\State Compute $\phi_{l_m''}$ $\forall l_m''\in\mathcal{L}''$ with \eqref{PsiEq}
\State Compute $R_m(l_m'')$ $\forall l_m''\in\mathcal{L}''$ with \eqref{rate_eq}
\end{algorithmic}
\end{algorithm}
\subsection{Benchamrk}
We consider the rate maximisation problem with outage constraints in \cite{kallehauge2022predictive} as the benchmark. The authors proposed a quantile-prediction-based scheme using Gaussian processes for the radio map construction. They departed from the estimation of the $\zeta-$quantile $q_{\zeta}(l_m)$ of $\ln(\Upsilon(l_m))$ at each observed location $l_m$ as the $\lfloor N\zeta\rfloor$-th order statistic. Then, they follow the procedure described in steps $(4)-(8)$ of \textbf{Algorithm 1} for obtaining the denormalised predictive mean and covariance matrices of the $\zeta-$quantile at the unobserved locations (same procedure used for $\hat{\mu}(l_{m''})$ in these steps). Finally, they proposed the following rate selection function 
\begin{equation}\label{rate_framework}
 R_m(l_{m''}) = \log_2\Big(1 + \exp\big( \theta_{l_{m''}}+\sqrt{2}\alpha_{l_{m''}}\text{erf}^{-1}(2\delta-1) \big)\Big),   
\end{equation}
where $\theta_{l_{m''}}$ represents the predictive mean of the $\zeta-$quantile of the SNR at location $l_{m''}$ and $\alpha_{l_{m''}}$ the corresponding standard deviation. Moreover, $\text{erf}^{-1}(\cdot)$ represents the inverse error function, and $\delta$ comes from applying the principle of meta probability which ensures that the outage requirements are met for any deployment of the observation points with probability $1-\delta$.

While this approach focuses the analysis on the distribution tail, it does not capture its heaviness, tail decay rate, and data dispersion, which are vital for efficiently modelling the URLLC region. Moreover, this method explicitly needs $N\ge1/\zeta$, which becomes prohibitive when $\zeta$ is extremely small. As discussed in the following section, these aspects are handled more efficiently by the EVT-based approach.

\section{Numerical Results}\label{section_4}
To evaluate the performance of the proposed method, we consider that the BS is providing service to a 
100 m $\times$ 100 m 
area. This region is divided into a regular grid of $120\times120$ (14400) different locations, thus, implying a vertical/horizontal spacing of 0.83~m. The BS knows $N$ SNR measurements at $M = 500$ locations randomly selected from the grid.
The environment and channel coefficients $h$ are simulated using QuaDRiGa v2.6.1 with the scenario \text{``3GPP\_3D\_UMi\_LOS"} (3GPP 3D Urban Micro-cell with line-of-sight)\cite{quadriga}. The channel realisations at each location are obtained by performing uniform variations to the phase of each cluster of the geometric channels provided by QuaDRiGa \cite{995515}. Besides the SNR measurement at the $N$ locations, we also have test data $(\Upsilon(l_{m''}),\hat{\xi}(l_{m''}),\hat{\sigma}(l_{m''}) \ \text{and} \ \hat{\mu}(l_{m''}))$ for each point $l_{m''}$ in the grid. The test data validates the algorithm when predicting $R_m$ at each location. The operation frequency is set to 1.5~GHz,  the transmit power $p$ at the BS to 1~mW, and the BS height is 10~m. Moreover, the noise power is given by $\upsilon^2 = -173.8+10\log_{10}BW+ NF$, where $NF$ is the noise figure and  $BW$ is the bandwidth. We set $NF=$7~dB and  $BW=100$ KHz, and adopt $f = -\ln(\cdot)$. All the simulation parameters are displayed in Table \ref{table_1}. 
\begin{table}[t!]
    \centering
    \caption{Simulation parameters}
    \label{table_1}
    \begin{tabular}{c c}
        \hline
        \textbf{Parameter} & \textbf{Value} \\
            \hline              $p$ & 1 mW \\ 
            $N$  & $10^3-10^6$ \\
            $NF$ &  7 dB\\
            $BW$ & $100$ kHz \\
            Frequency & 1.5 GHz \\
            $\tau$ &  $10^{-5}-10^{-3}$\\
            $\rho$ & $0.99$ \cite{dumouchel1983estimating}\\ 
             $\delta$ & $10^{-3}$ \cite{kallehauge2022predictive}\\ 
         $\zeta$ &  $10^{-5}-10^{-3}$ \\  
          $M$ &  $500$ \\ 
           BS height &  10 m \cite{quadriga}\\  
            UE height &  1.5 m \cite{quadriga}\\ number of multipath  clusters & 10\\ 
            \hline
    \end{tabular}
\end{table}

\begin{figure}[t!]
    \centering
    \includegraphics[width = \columnwidth,height = .5\columnwidth]{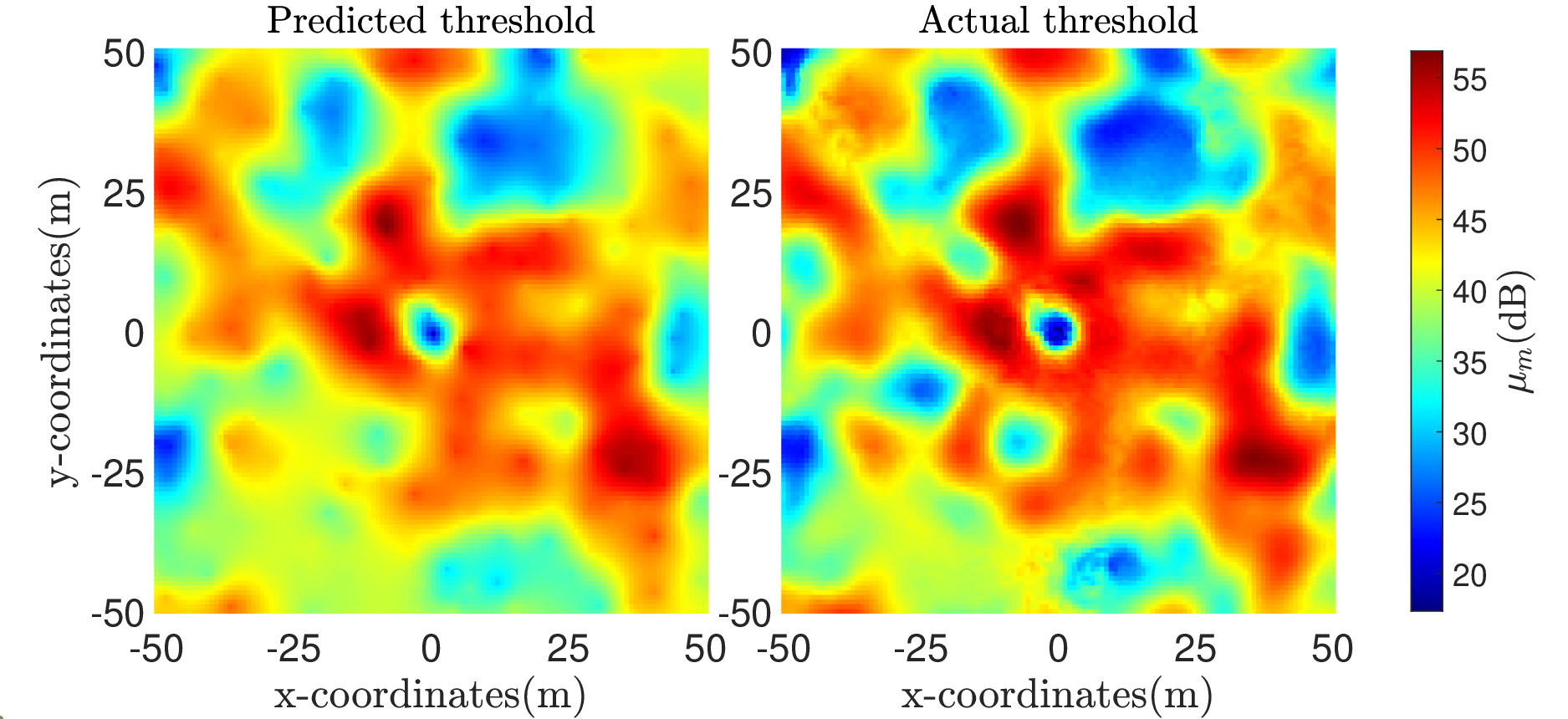}
    \caption{Predictive mean of the threshold $\hat{\mu}$ (left) at each location in the coverage area, and actual threshold $\mu$ (right) obtained from the test data. The number of samples used is $N=10^5$ and $\rho = 0.99$.}
    \label{threshold}
\end{figure}

Fig. \ref{threshold} shows the predictive mean of the threshold values $\hat{\mu}$ of the GPD at each location in the coverage area and the actual threshold values ${\mu}$  obtained from the test data. Notice that the predictive mean is generally close to the actual parameters in most grid points, validating the adopted predictive approach. However, since the range of possible values for $\hat{\mu}$ in log-scale is large, relatively small variations in the predictions can indeed represent significant deviations from the actual value. Thus, it is recommended to consider the $\tau-$quantile of $\hat{\mu}$ so the proposed algorithm does not overestimate the channel conditions at any location.

\begin{figure}[t!]
    \centering
    \includegraphics[width = \columnwidth,height = .5\columnwidth]{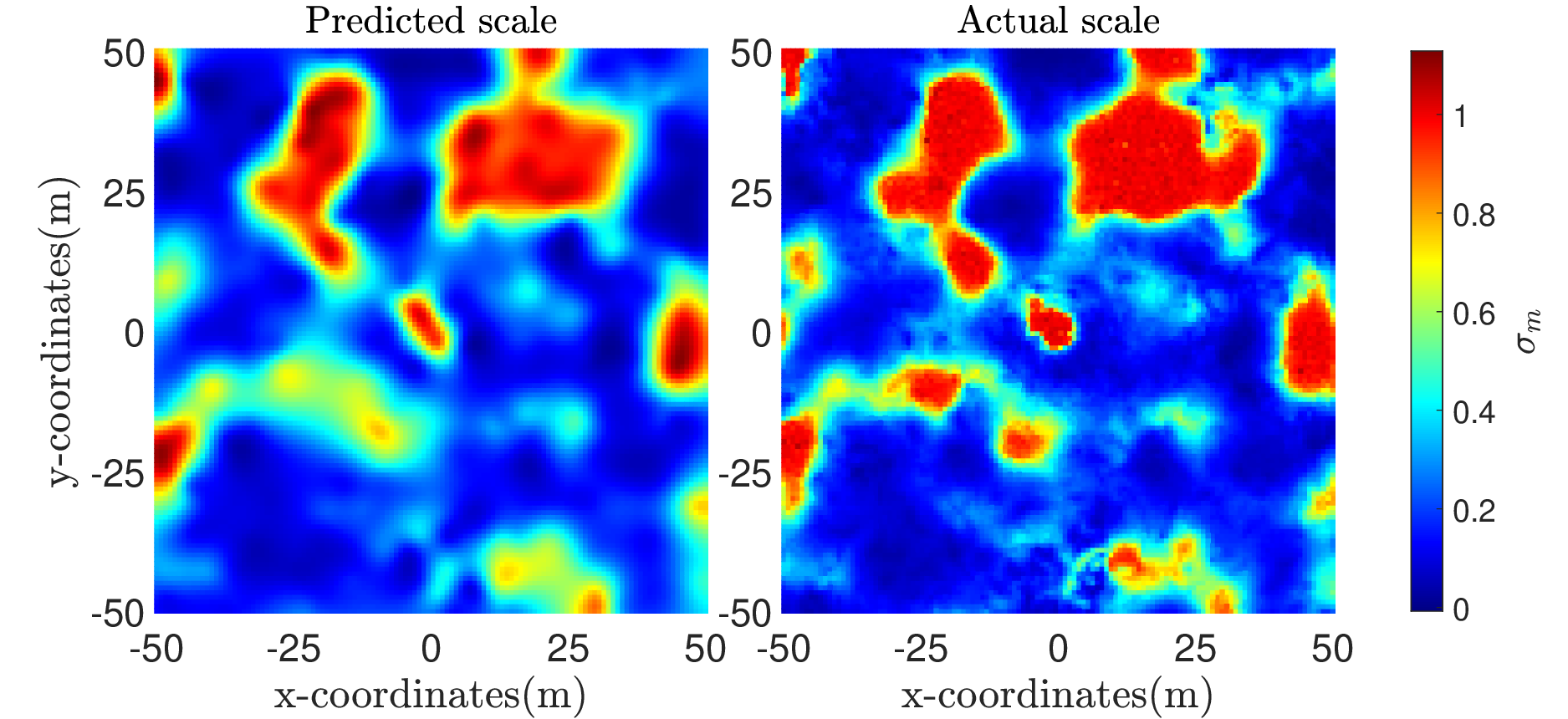}
    \caption{Predictive mean of the scale $\hat{\sigma}$ (left) at each location in the coverage area, and actual scale $\sigma$ (right) obtained from the test data. The number of samples used is $N=10^5$ and $\rho = 0.99$.}
    \label{scale}
\end{figure}
\begin{figure}[t!]
    \centering
    \includegraphics[width = \columnwidth,height = .5\columnwidth]{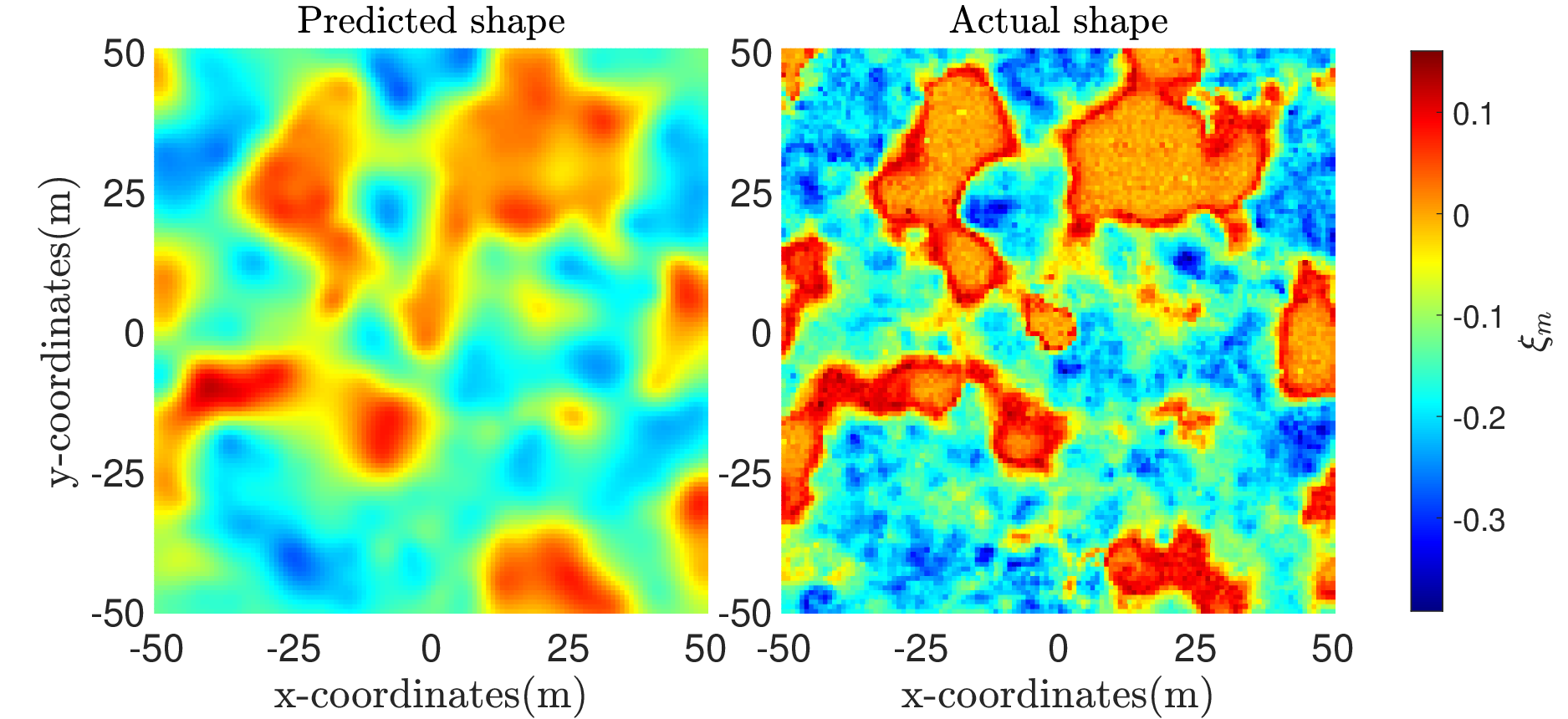}
    \caption{Predictive mean of the shape $\hat{\xi}$ (left) at each location in the coverage area, and actual shape $\xi$ (right) obtained from the test data. The number of samples used is $N=10^5$ and $\rho = 0.99$.}
    \label{shape}
\end{figure}

Fig. \ref{scale} displays the predictive mean of the scale $\hat{\sigma}$ alongside the actual scale $\sigma$, while Fig. \ref{shape} shows the predictive mean of the shape $\hat{\xi}$ and actual shape $\xi$ parameter. Notice that the predictions are also close to the actual values, validating both the general framework and the suitability of the Matérn correlation model for this scenario. Interestingly, the areas with higher values in these figures tend to overlap with those with lower values in Fig. \ref{threshold}. This is related to worse channel conditions or a distribution tail of the SNR that is less heavy or has more data dispersion. Also, inserting quantiles to these parameters to handle variations in the predictions will cause no/little variations in the contours of the distribution tail. This is because the range of predicted values is considerably small, and the variances of the estimates are $\ll 1$. This is why the predictive means of scale and shape parameters are adopted as the input for \eqref{Outage_eq_3} as we analyse next.

\begin{figure}[t!]
    \centering
    \includegraphics[width = \columnwidth,height = .5\columnwidth]{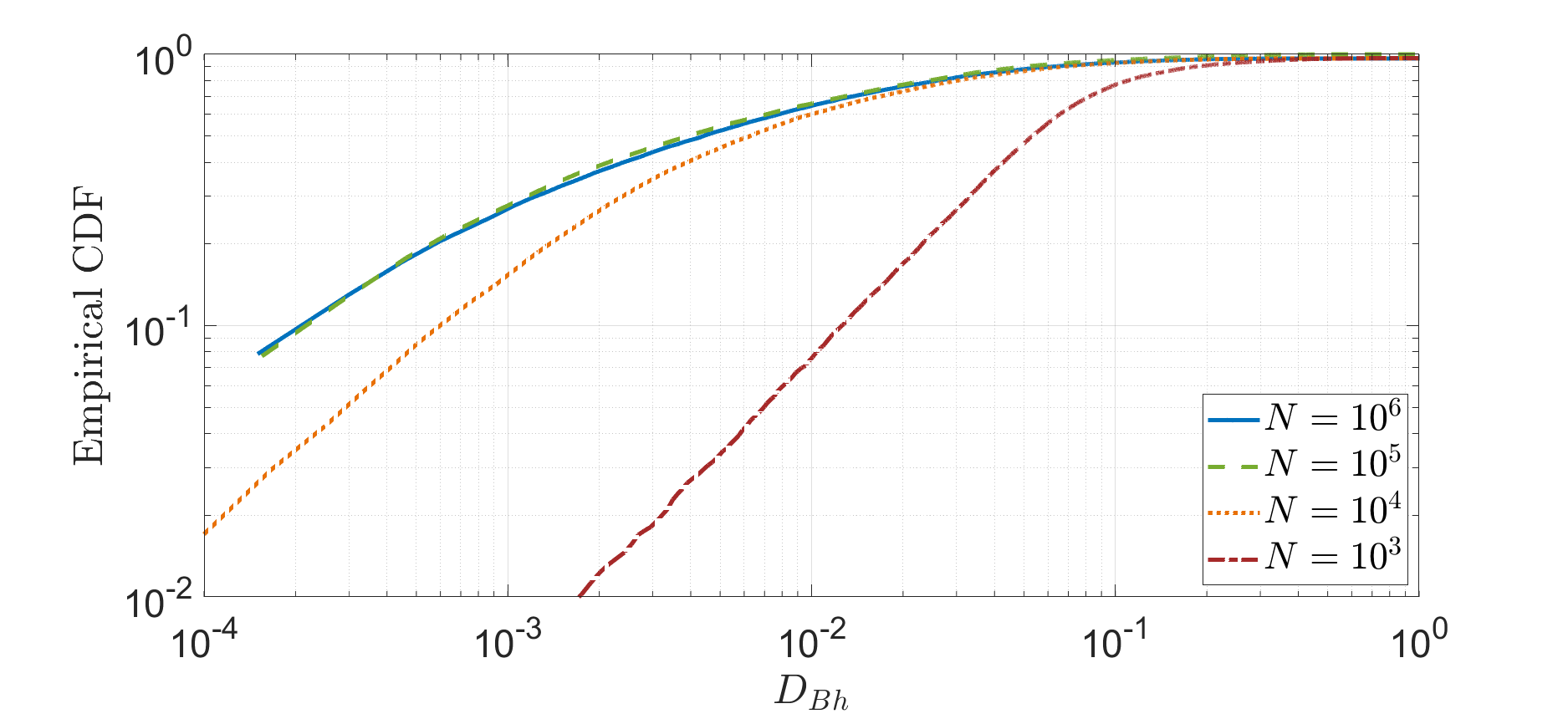}
    \caption{Empirical CDF of the Bhattacharyya distance between predicted GPD and actual GPD with a common threshold value for a different number of samples with $\rho=0.99$.}
    \label{distance}
\end{figure}

Fig. \ref{distance} analyses the accuracy in the predictions of scale and shape parameters of the GPD and the impact of the number of samples. For the analysis, a common threshold value was used ($\mu = 0$), and the Bhattacharyya distance ($D_{Bh}$) was considered as the metric to measure divergence \cite{bhattacharyya1943measure}. This distance is defined as 
\begin{equation}\label{Bdistance}
    D_{Bh} = -\ln\Big(\int_\mathcal{W}\sqrt{g_1(w)g_2(w)}dw\Big),\nonumber
\end{equation}
where $\mathcal{W}$ depicts the variable domain and $g_1(\cdot)$ and $g_2(\cdot)$ represent the compared probability density functions. The distance is $D_{Bh}\ge 0$, where $D_{Bh} = 0$ corresponds to two identical distributions. The figure shows the empirical CDF of $D_{Bh}$ obtained for each point in the grid. Notice that for $10^4\le N\le10^6$, nearly $100\%$ of the obtained distances are below $10^{-1}$, which denotes a high accuracy in predicting scale and shape parameters. Moreover, larger samples $N$ improve the prediction since the estimated scale and shape at the $M$ locations get closer to the actual values. However, this improvement decreases as $N \rightarrow 10^3\rightarrow 10^4\rightarrow 10^5\rightarrow 10^6$, being almost not noticeable when $10^5\rightarrow 10^6$, which means that the estimates are extremely close to the actual parameters when $N = 10^5$ for $\rho = 0.99$. 

\begin{figure}[t!]
    \centering
    \includegraphics[width = \columnwidth,height = .5\columnwidth]   {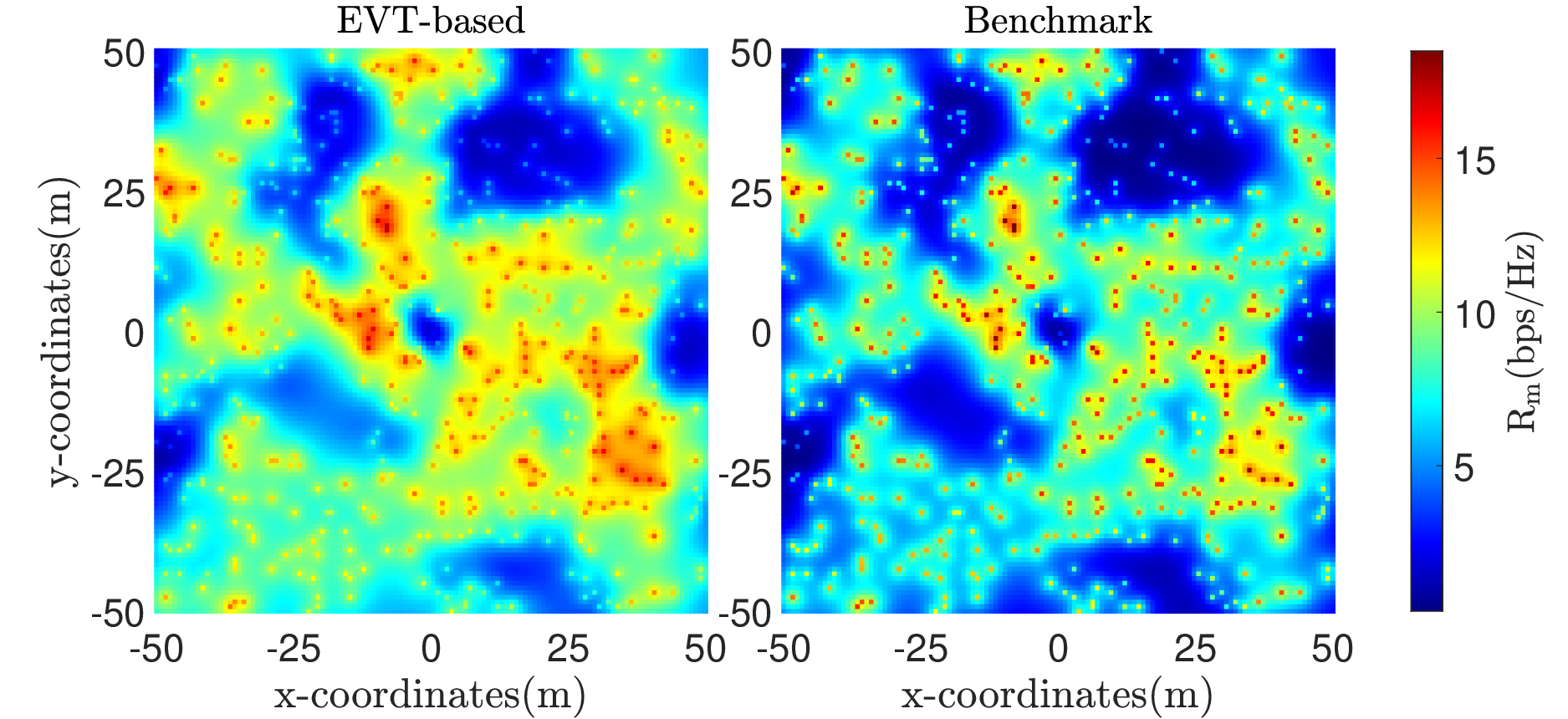}
    
    \caption{Rate selection at each location in the coverage region for the EVT-based approach (left) and the benchmark (right). The target outage probability is $\zeta = 10^{-3}$ with  $N=10^5$ and $\rho = 0.99$.}
    \label{rate_comparisson}
\end{figure}

\begin{figure}[t!]
    \centering
    \includegraphics[width = \columnwidth,height = .5\columnwidth]   {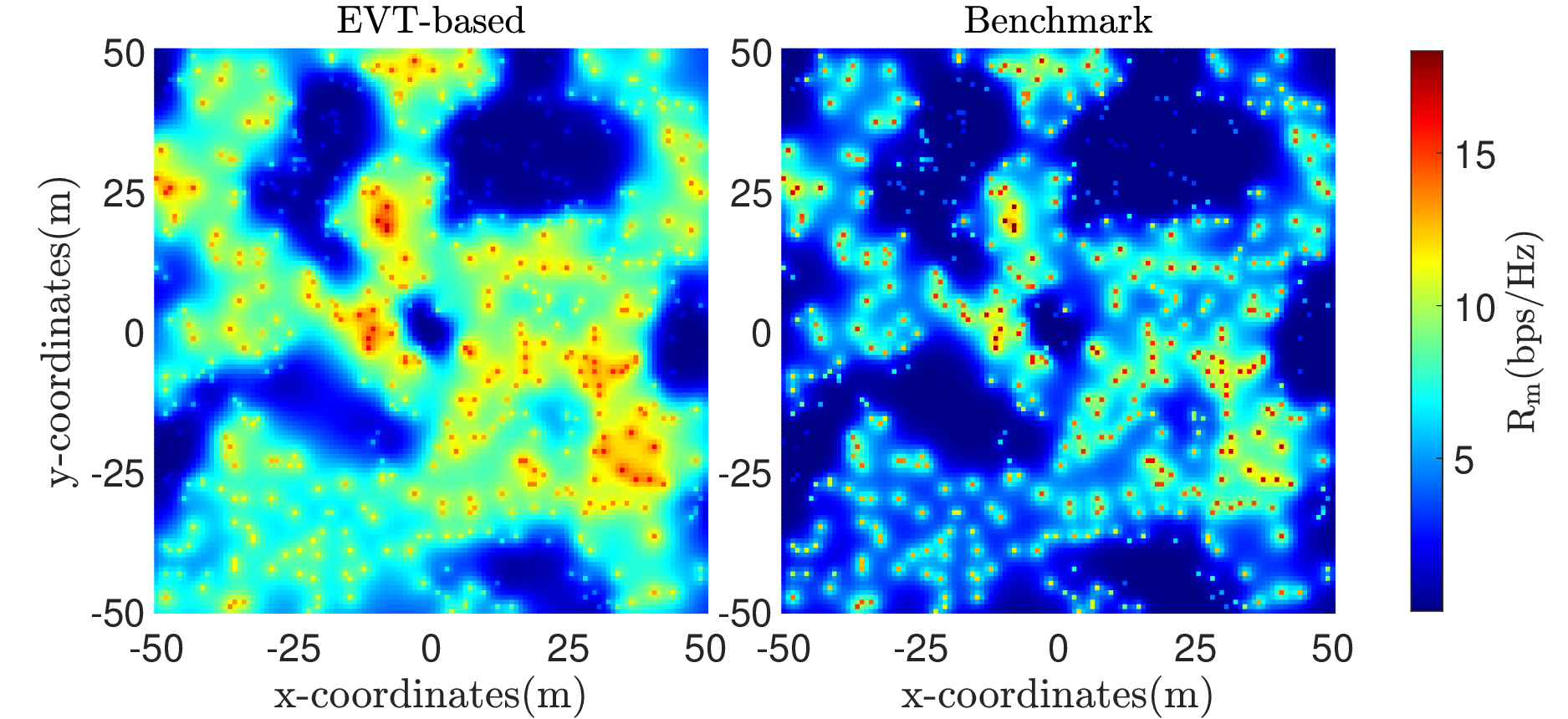}
    
    \caption{Rate selection at each location in the coverage region for the EVT-based approach (left) and the benchmark (right). The target outage probability is $\zeta = 10^{-4}$ with  $N=10^5$ and $\rho = 0.99$.}
    \label{rate_comparisson2}
\end{figure}

Fig. \ref{rate_comparisson} and Fig. \ref{rate_comparisson2} show the rate allocation when using the proposed EVT-based approach and the benchmark for $\zeta= 10^{-3}$ and $\zeta= 10^{-4}$, respectively. The figures clearly show that the EVT approach outperforms the benchmark for both outage targets in the coverage area. This means that our presented scheme predicts tail behaviour more accurately, thus performing a more efficient rate selection while guaranteeing outage demands, as we will discuss next. Computing the mean of the predicted rates in the coverage area, we have that the EVT-based approach transmits at a rate  $28.1\%$ higher than the benchmark for $\zeta = 10^{-3}$, and $65.7\%$ higher for $\zeta = 10^{-4}$.

\begin{figure}[t!]
    \centering
    \includegraphics[width = \columnwidth,height = .5\columnwidth]{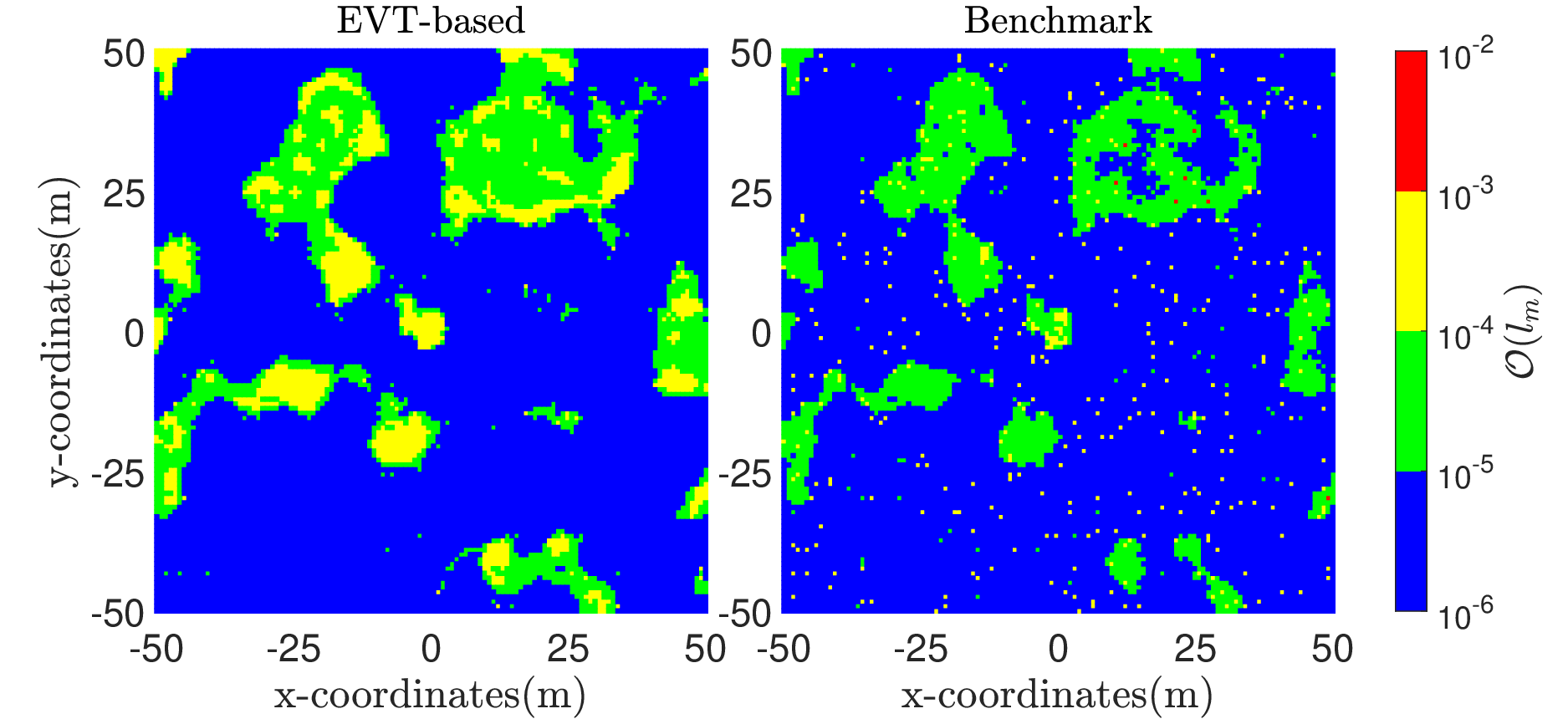}
    
    \caption{Achieved outage probabilities at each location in the coverage region for the EVT-based approach (left) and the benchmark (right). The target outage probability is $\zeta = 10^{-3}$ with  $N=10^5$ and $\rho = 0.99$.}
    \label{out_comparisson}
\end{figure}

\begin{figure}[t!]
    \centering
    \includegraphics[width = \columnwidth,height = .5\columnwidth]{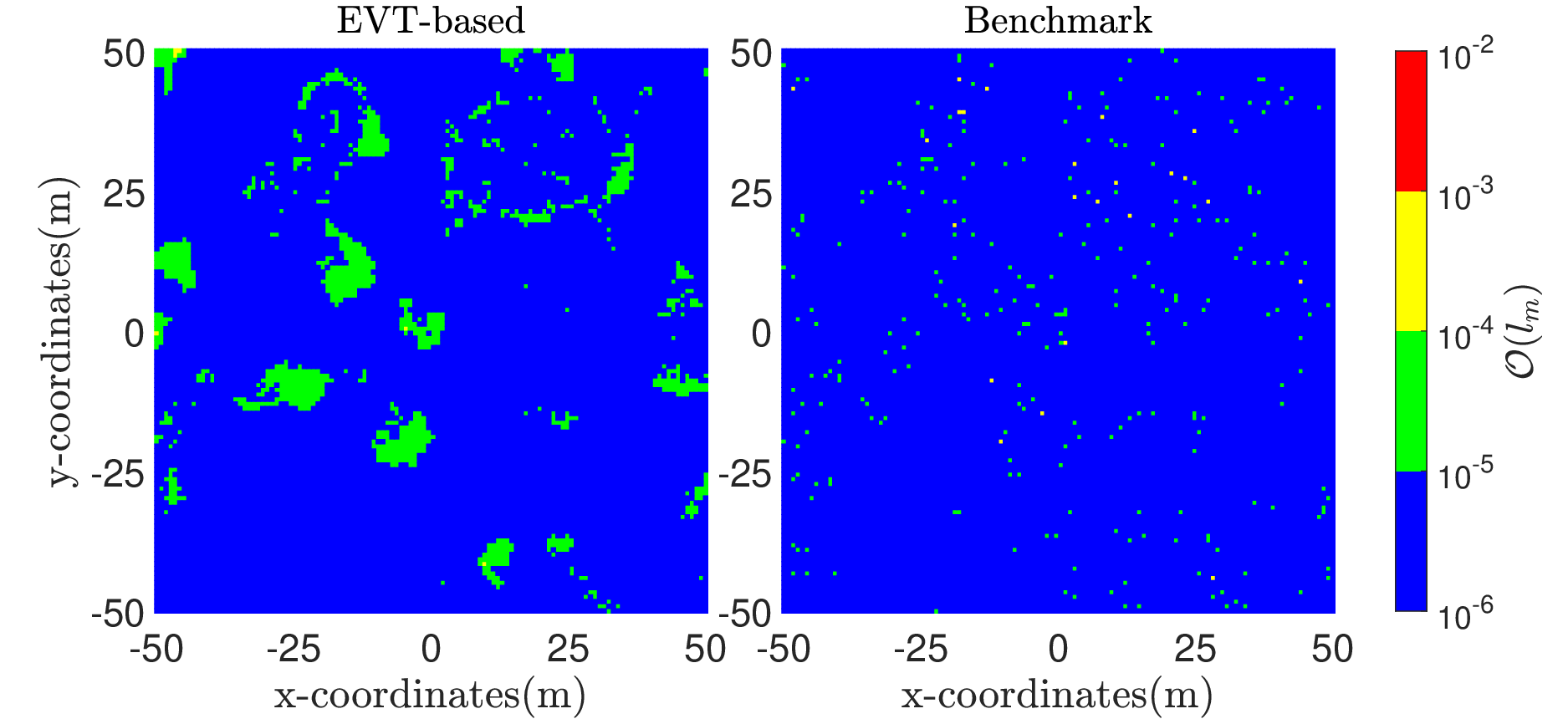}
    
    \caption{Achieved outage probabilities at each location in the coverage region for the EVT-based approach (left) and the benchmark (right). The target outage probability is $\zeta = 10^{-4}$ with  $N=10^5$ and $\rho = 0.99$.}
    \label{out_comparisson2}
\end{figure}
 Fig. \ref{out_comparisson} and Fig. \ref{out_comparisson2} display the achieved outage probabilities at each spatial position for targets $\zeta = 10^{-3}$ and $\zeta = 10^{-4}$, respectively. Note that the EVT-based approach consistently achieves outage probabilities closer to the target in a larger number of points in the area when compared to the benchmark. Also, the percentage of points where outage requirements are met (defined as availability hereinafter) is more prominent for our proposed scheme. 
\begin{figure}[t!]
    \centering
    \includegraphics[width = \columnwidth,height = .5\columnwidth]{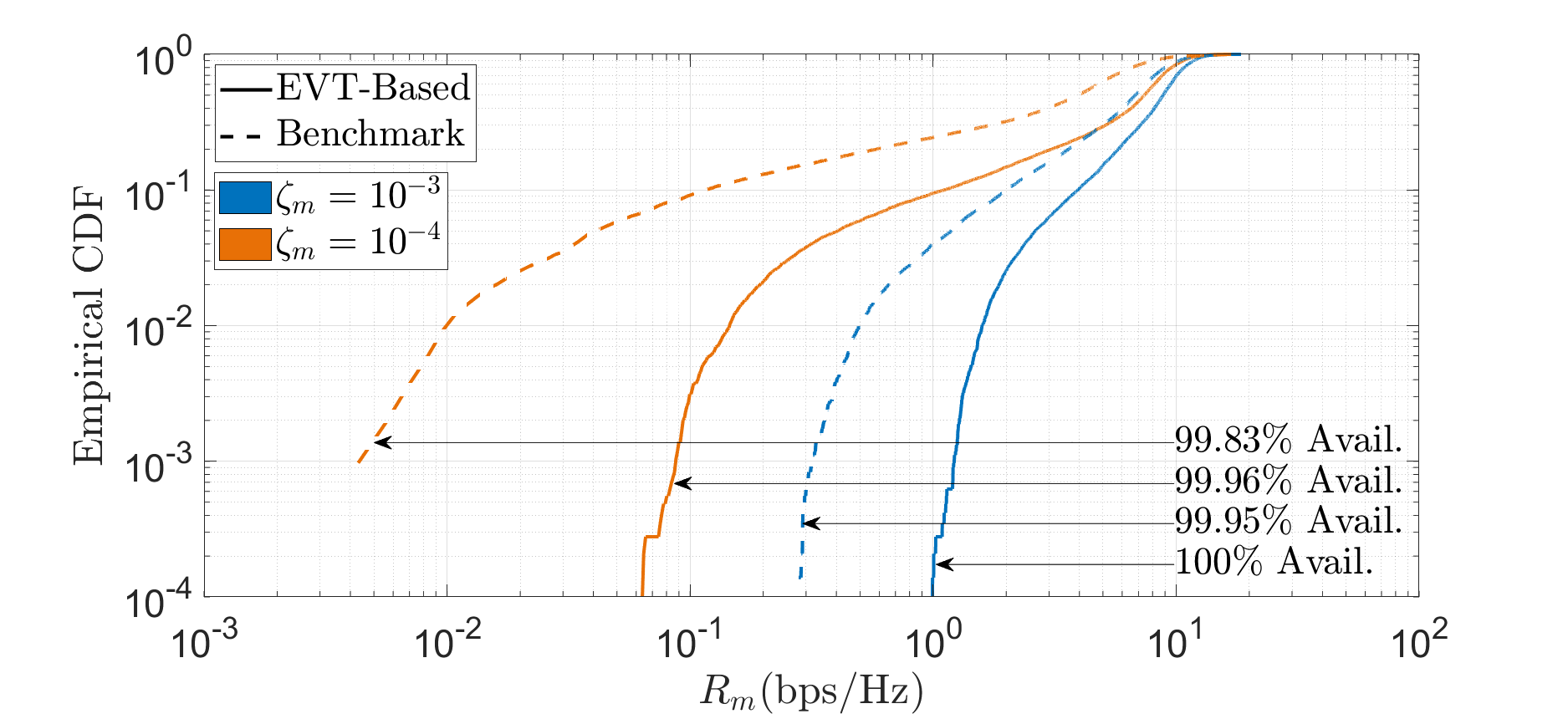}
    
    \caption{Empirical CDF of the rate for the EVT-based approach and benchmark. The target outage is set to $\zeta= [10^{-3}, 
    10^{-4}]$, with $N = 10^5$ and $\rho=0.99$.}
    \label{CDF_rate}
\end{figure}

Fig. \ref{CDF_rate} shows the empirical CDF of the achievable rate in the coverage areas and the associated availability for $\zeta=[10^{-3}, 10^{-4}]$. Notice that the performance gap is significant and larger as the outage probability requirements get stricter. The EVT approach has an availability of $100\%$ for $\zeta = 10^{-3}$ and $99.95\%$ for $\zeta = 10^{-4}$, superior to the benchmark in both cases.

\begin{figure}[t!]
    \centering
    \includegraphics[width = \columnwidth,height = .5\columnwidth]{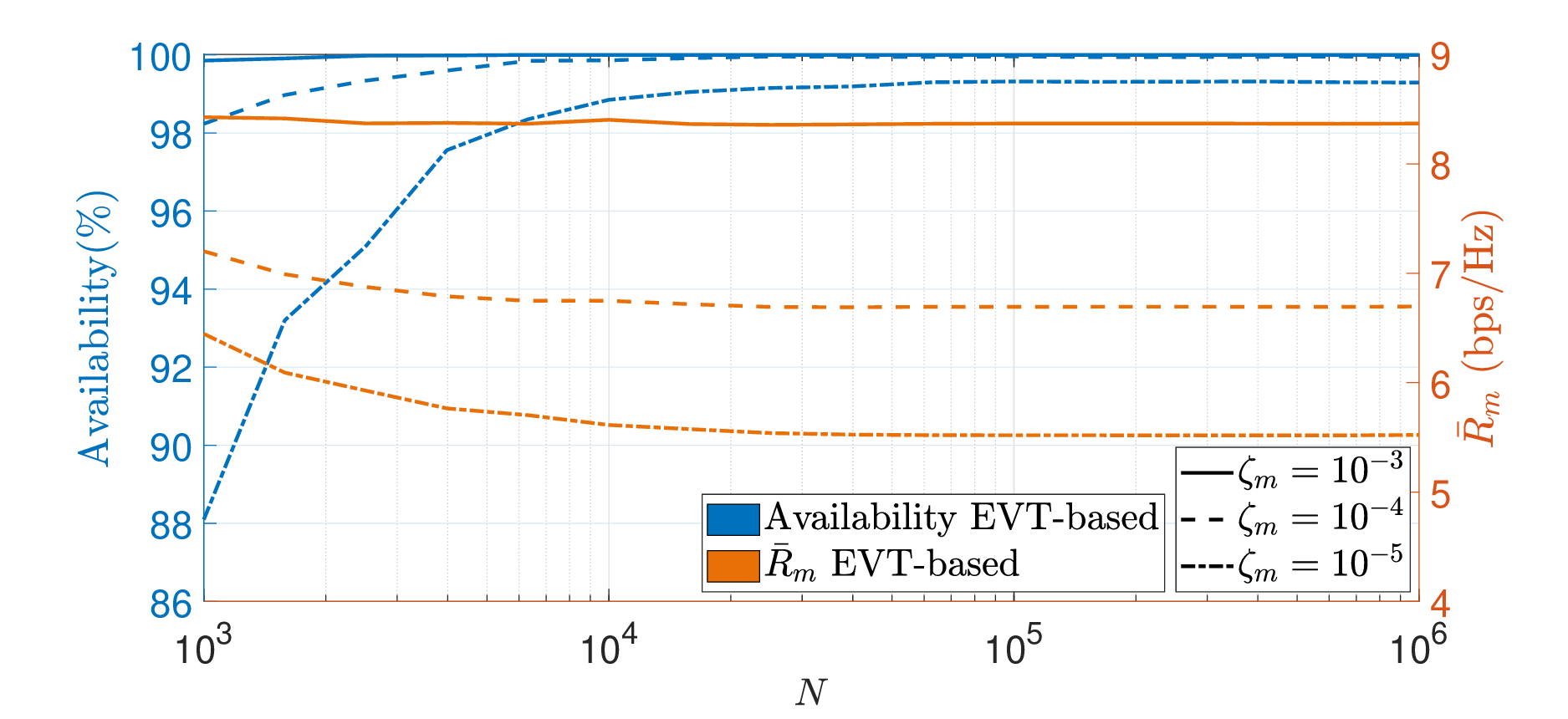}
    \caption{Availability and mean predicted rate for the EVT-based approach. The outage target is set to $\zeta = \{10^{-3}, 10^{-4},10^{-5}\}$ and $\rho = 0.99$.} 
    \label{OCA_rate}
\end{figure}

Fig. \ref{OCA_rate} shows the availability and mean predicted rate $(\bar{R}_m)$ for the EVT-based scheme. Notice that the availability is larger for a larger number of samples $N$ and for less strict outage targets. This is because the estimated parameters (scale and shape) get closer to the actual values as $N$ grows, which implies more accurate modelling of the tail region and is reflected in the spatial predictions. Notice that the rate $\bar{R}_m$ decreases with $N$ and tends to remain constant beyond a given $N$ where there are enough samples to perform an accurate fitting of the GPD for a constant $\rho$ as discussed in Fig.~ \ref{distance}. At this point, the availability also reaches a maximum which is bounded by the error in the spatial predictions. The figure also shows one key advantage of EVT which is the need for fewer samples when compared with most approaches in the literature. For instance, the $\zeta = 10^{-5}$ requirement is available in nearly $99\%$ of the coverage area when $N = 10^4$ samples are used, which is not feasible for the benchmark due to the need for $N>10^5$.

\section{Conclusions}\label{section_5}
In this research, we successfully demonstrated the effectiveness of integrating EVT with radio maps to enhance the reliability and performance of URLLC. Our innovative approach, which predicts the parameters of the GPD at unobserved locations using Gaussian processes, offers a flexible and robust solution for modelling extreme channel conditions. This method is particularly advantageous in scenarios requiring precise tail characterisation of the SNR distribution, which is critical for URLLC's requirements. A significant advantage of our proposed framework is its ability to achieve good performance with fewer samples, thus minimising the need for extensive data collection. This reduces the computational burden and enhances the model's applicability in dynamic and resource-constrained environments. Through a comprehensive evaluation focused on a rate maximisation problem with stringent outage constraints, our findings revealed that the proposed method consistently outperforms the benchmark. It guarantees outage demands across a more extensive portion of the coverage area and achieves higher transmission rates.

\bibliographystyle{IEEEtran}
\bibliography{IEEEabrv,bibliography}
\end{document}